\documentclass[grl]{agu2001}
\usepackage{lineno}
\linenumbers*[1]

\usepackage{amsmath,textcomp}
\usepackage{times}
\usepackage{graphicx}

\def\m1{{$^{-1}$}}

\def\beq{\begin{equation}}
\def\eeq{\end{equation}}

\authorrunninghead{SOBEL ET AL.}

\titlerunninghead{MULTIPLE EQUILIBRIA}

\begin{document}

\title{Multiple Equilibria in a Single-Column Model of the Tropical Atmosphere}

\author{A. H. Sobel}
\affil{Department of Applied Physics and Applied Mathematics
and Department of Earth and Environmental Sciences,
Columbia University, New York, NY, USA}

\author{G. Bellon}
\affil{Department of Applied Physics and Applied Mathematics,
Columbia University, New York, NY, USA}

\author{J. Bacmeister}
\affil{Goddard Earth Sciences and Technology Center, University of Maryland, Baltimore County, 
Baltimore, MD, USA}

\begin{abstract}

A single-column model run under the weak temperature gradient approximation,
a parameterization of large-scale dynamics appropriate for the tropical atmosphere,
is shown to have multiple stable equilibria.  Under conditions permitting persistent deep
convection, the model has a statistically steady state in which such convection occurs,
as well as an extremely dry state in which convection does not occur.  Which state is
reached depends on the initial moisture profile.  

\end{abstract}

\begin{article}
\newpage
\section{Introduction}

In the extratropics, balanced dynamics
associated with baroclinic eddies can force precipitation strongly by inducing
ascent and adiabatic cooling.  In the tropics, the reasons for the onset
of precipitation at a given time and place
are often much more subtle, and large-scale ascent 
more a response to deep convection than a cause.  On the other hand, in the
tropics the relationship of time-mean precipitation to boundary conditions
is generally stronger than it is in the extratropics.  In simple models of
tropical climate dynamics, it is typical to treat the time-mean tropical precipitation
as deterministically related to the boundary conditions.  While this may for many purposes be 
adequate, it is clear that the time mean state, particularly in climatologically 
rainy regions, is an average over periods of strong precipitation and 
periods of negligible precipitation.  As precipitation cannot be negative, 
this implies an inherent nonlinearity that is 
overlooked in deterministic models of the time mean state.  It also 
suggests that transient disturbances, or other factors whose relationship to 
the boundary conditions is at best hard to discern, may play a role in 
maintaining the time-mean state. 

In this study we present a very simple expression of these properties of the 
tropical atmosphere, namely its nonlinearity and the potential
complexities hidden in its time-mean response
to boundary conditions.  We show that a single-column model with boundary conditions
permitting strong deep convection can have two stable steady (or statistically steady)
states:  one in which persistent deep convection occurs, and one in which it does
not.  Which solution is reached in numerical integrations depends on 
the initial moisture field.  

\section{Model and Experiment Design}

We use the single column version of the GEOS-5 general circulation model (GCM).
A description of the GEOS-5 system and its physical parameterizations may be found at 
http://gmao.gsfc.nasa.gov/systems/geos5/, and a brief outline is presented in the 
auxiliary materials.

The model is run over an ocean surface with
fixed SST.  The insolation and
solar zenith angle are held constant at values of 400 $W~m^{-2}$ and zero 
respectively.  The surface wind speed, used in the bulk formulae for the
surface fluxes, is set to a constant $7~m~s^{-1}$.

The experiment design is essentially the same as that used in \nocite{SB00} {\textit Sobel and Bretherton} [2000;  
hereafter SB00].  The model is first run to a state of radiative-convective equilibrium 
(RCE) over an SST of 301 K.
In this calculation the large-scale vertical velocity is set to zero.  The temperature profile 
from the RCE calculation is
then used as an input to a set of calculations in which the model is modified
to implement the weak temperature gradient (WTG) approximation.  In the WTG
calculations, the temperature profile is held fixed in time in the free troposphere, 
defined somewhat arbitrarily as those levels with pressures less than 
$p_{\rm inv}$= 850 hPa.  At those levels, the large-scale
vertical velocity is diagnosed as that which
causes the vertical advection of potential temperature to precisely balance
the diabatic heating computed by the model physics, consistent with the requirement
of zero temperature tendency.  Horizontal advection of temperature is also assumed
negligible.  In the nominal boundary layer, defined as levels with pressures greater
than $p_{\rm inv}$, the temperature is determined prognostically, with the vertical velocity
computed by linear interpolation in pressure between the diagnosed value at $p_{\rm inv}$
and an assumed value of zero at the surface.  At all levels, large-scale
vertical advection of
humidity and condensate --- both of which are integrated prognostically as
usual --- are computed using the vertical gradients derived from
the internally predicted profiles of
those variables and the large-scale vertical velocity diagnosed as described
above.

In some experiments, the initial moisture profile is taken from the steady-state
RCE solution.  In others, the initial moisture profile is obtained by setting the 
free-tropospheric humidity to zero.  

In some experiments, horizontal advection of moisture is neglected, as in SB00.  In
others, it is parameterized by a relaxation back to a fixed profile - the steady-state RCE solution - with a fixed time scale.  The time scale can be thought of as an advective one given by a 
length scale associated with horizontal moisture gradients divided by a velocity,
where the velocity is thought of as a rotational one which is independent of height
(so that the time scale itself is independent of height),
and also independent of the magnitude of the
divergent circulation implied by the vertical velocity in the column.  Another
approach is to consider advection by that divergent circulation itself, in which
case the relaxation rate is simply $d\omega/dp$, where $\omega$ is the pressure
vertical velocity and $p$ the pressure \citep{RZ05}.  We have performed
calculations with this method also;  the results are not qualitatively different 
from those using the 
fixed advective time scale.

All simulations are run for one year, by which time the model
has reached a solution that is statistically steady, and in many cases 
close to truly steady;  oscillations sometimes occur, as described 
briefly below.  Results
presented are averages over the last 2 months.

\section{Results}

\subsection{GEOS 5 SCM}

\begin{figure}
\noindent
\includegraphics[height=6cm,clip]{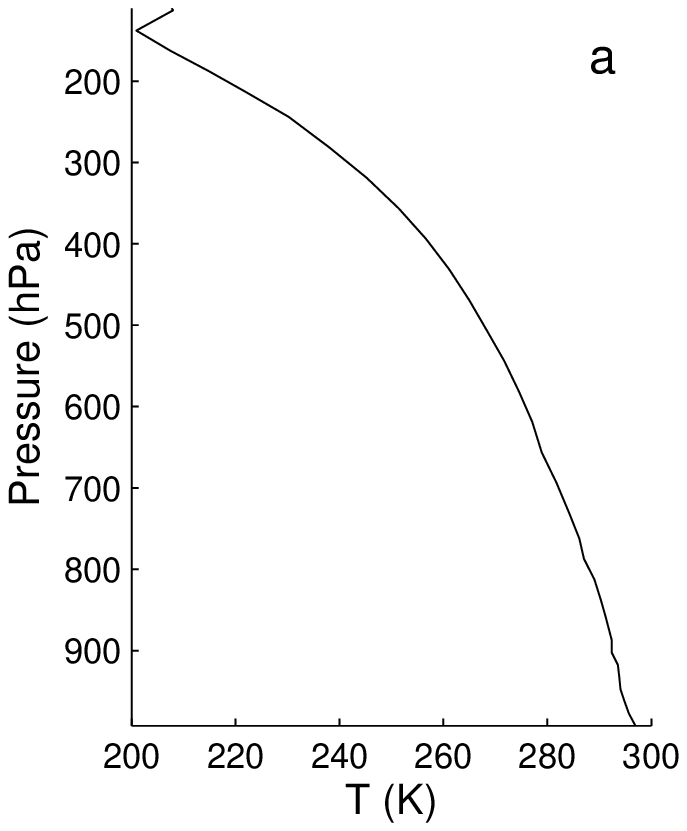}
\includegraphics[height=6cm,clip]{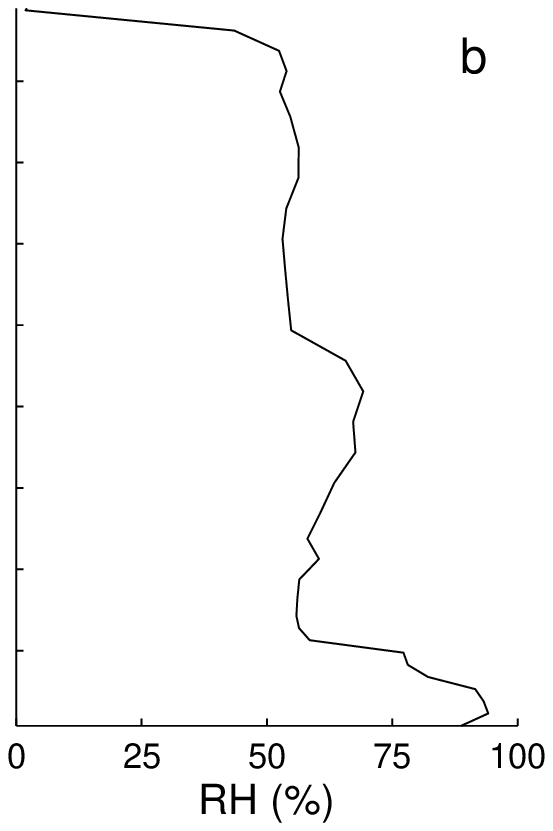}
\caption{Temperature (a) and relative humidity (b)
 profiles in the radiative-convective
equilibrium state over an SST of 301K.  Temperature is in $^\circ K$,
relative humidity in \%.}
\label{fig:rce}
\end{figure}

Fig.~\ref{fig:rce} shows the temperature and relative humidity profiles obtained from the RCE
calculation.  The temperature profile is close to moist adiabatic as expected.
Fig. 2 shows time-mean precipitation as a function of SST for the 
WTG experiments.  The figure is similar to fig. 4 of SB00;
each point represents a different experiment, each of which uses the same free-tropospheric
temperature profile, obtained from the RCE calculation.  The difference here [besides
that we use a different model;  SB00 used the model of \cite{Renno94a}] is that 
we show several sets of curves, obtained using different initial moisture profiles, and
different time scales for horizontal moisture advection (SB00 did not include this
process in any of their calculations). 
We first describe the solid curves, which were obtained using no horizontal 
moisture advection.  

When 
the initial moisture profile is taken from the RCE, a rainy state is obtained for
sufficiently large SST.  The shape of the curve of precipitation vs. SST for this
set of solutions (the upper blue curve) is qualitatively similar to that obtained by SB00.  The 
lower blue curve curve shows solutions obtained using an initial moisture profile is set to 
zero in the free troposphere.

When the initial profile is sufficiently dry, apparently a dry state can be maintained
although the SST is high and the convective available potential energy (CAPE) 
for an undilute parcel ascent would be substantial.
For a dry profile, the convective parameterization is unable to generate
significant convective heating or precipitation, presumably due to inhibition by entrainment of
the dry air.  In the absence of convective heating, radiative cooling forces descent,
which 
maintains the dry state in the free troposphere.  On the other hand if the 
initial conditions are sufficiently moist, for sufficiently large SST the deep convective
scheme is able to become and remain active.  If the ensuing heating exceeds the radiative cooling, it 
induces large-scale ascent, which moistens the atmosphere further by 
large-scale advection, leading to the maintenance of the convective state.
For SST below a critical value, here around 300.5 K, there is insufficient CAPE
(or strictly, work function;  Arakawa and Schubert 1974) for deep convection to occur
even if the initial profile is moist.  Dry air can suppress deep convection in the
presence of large CAPE, but moist air cannot cause it to occur if CAPE is absent.
Thus the multiple equilibria occur only for SST above the critical value.

\begin{figure}
\noindent
\includegraphics[height=8cm,clip]{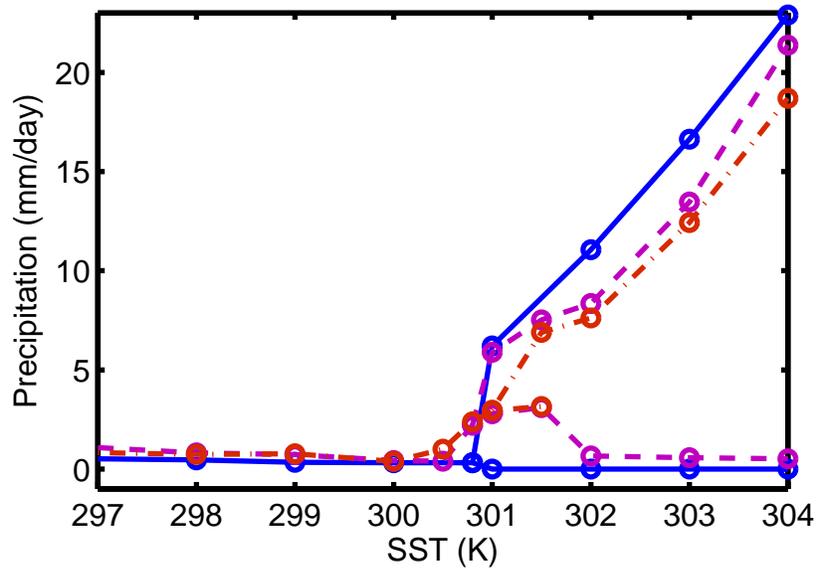}
\caption{Precipitation ($mm~d^{-1}$) as a function of SST in the convective and
non-convective equilibria, with no horizontal moisture advection (blue, solid) and
with horizontal moisture advection parameterized as a relaxation
back to the RCE profile with a time scale of 6 days (magenta, dashed) and 3 days (red, dot-dashed).
Each circle represents a separate model integration.}
\label{fig:p_sst}
\end{figure}

Fig. 3 shows vertical profiles of relative humidity as functions of SST in the dry state and the
moist state, without horizontal moisture advection.  
We see that the relative humidity in the boundary layer is
similar in both sets of solutions, but that the free-tropospheric humidity
is essentially 
zero in the dry case;  in the absence of horizontal moisture advection, there
is no source of moisture to balance drying due to descent.  This is somewhat 
artificial, as in the real atmosphere even the driest desert is connected to 
regions of finite humidity by horizontal advection.  The magenta and red
curves in fig.~\ref{fig:p_sst} show results using horizontal advective time scales
of 3 and 6 days, respectively (the no-advection case would correspond to an 
infinite advective time scale);  again the relaxation is towards a target
moisture profile equal to that in the RCE.  We see that for the 3-day time
scale, multiple equilibria exist only over a narrow range of SSTs.  The
horizontal advective moistening in this case prevents
the atmosphere from becoming sufficiently dry to inhibit deep convection
once SST exceeds the RCE value of 301 K by a small increment.  On the other
hand, for a mixing time scale of 6 days, the multiple equilibria persist 
up to large SST.  In this case, the horizontal advective moistening is 
too weak to overwhelm the vertical advective drying, and the atmosphere
can stay dry enough to inhibit deep convection even for large SST 
if it is initialized sufficiently dry.  These results show that the 
existence of multiple equilibria
is not purely an artifact of the neglect of horizontal moisture advection, although
sufficiently strong horizontal moisture advection can eliminate the dry 
equilibrium (or render it unstable so that it cannot be achieved numerically).

In the cases with horizontal moisture advection, the precipitation is a
non-monotonic function of SST, with local maxima near 301K, the value 
at which the RCE was computed.  The solutions in this neighborhood
are time-dependent, either periodic or chaotic but with a dominant spectral
peak.  The periods range from 1-20~$h$, and maximum excursions of
0.5-10~$mm~~d^{-1}$, with the largest oscillations tending to occur for the
higher SST values although the dependence of amplitude on SST is not
monotonic.  We do not understand these oscillations in detail and do not
address them further here;  our interest is in the presence or absence 
of multiple equilibria.

\subsection{Interpretation}
We consider under what conditions the multiple equilibria can exist.
We can derive some general constraints that any model must obey if it is
to obtain a dry solution under boundary conditions and forcings that also
allow a rainy solution.

Starting from the primitive equations,
the steady temperature and moisture equations in pressure coordinates are
\begin{eqnarray}
{\bf u} \cdot \nabla T+ 
 \omega \partial _p s  = Q_c + R - \partial_p \overline{\omega's'},
\label{eq:primeq_T}\\
{\bf u} \cdot \nabla q+ 
 \omega \partial _p q  = Q_q - \partial_p \overline{\omega'q'},
\label{eq:primeq_q}
\end{eqnarray}
where ${\bf u}$ is horizontal velocity, $T$ temperature in energy units (i.e., multiplied by the heat capacity of air at constant pressure $C_p$), $s$ dry static energy, $q$ specific humidity in energy units (i.e., multiplied by the latent heat of vaporization $L_v$), $R$ radiative heating, $Q_c$ convective heating, $Q_q$ convective moistening, $\omega$ large-scale vertical velocity, and $\nabla$ the horizontal gradient on pressure surfaces. 
$\overline{\omega's'}$ and $\overline{\omega'q'}$ are the turbulent fluxes, limited to the atmospheric boundary layer (ABL), as parameterized by the model's boundary layer scheme;  in the
free troposphere these fluxes are incorporated into $Q_c$ and $Q_q$.
  
The WTG approximation requires neglect of horizontal temperature advection in the free troposphere;
for the sake of argument let us also neglect horizontal moisture advection and
horizontal temperature advection in the ABL.
If there is no deep convection, $Q_c=Q_q=0$, the equations for the free troposphere are:
\begin{eqnarray}
\omega \partial _p s  = R ,
\label{eq:primeq_T'}\\
\omega \partial _p q  = 0 .
\label{eq:primeq_q'}
\end{eqnarray}
Therefore, unless the radiative cooling $R$ is zero, the vertical velocity $\omega$ is non-zero and $\partial _p q = 0$. Integrating from the top of the atmosphere, this in turn yields $q=0$ : the free troposphere has to be dry for the non-convective equilibrium to exist.  If horizontal moisture advection
is included, the free-tropospheric humidity can be non-zero in this dry state.

Equations (\ref{eq:primeq_T'}) and (\ref{eq:primeq_q'}) can be integrated over the ABL: 
\begin{eqnarray}
\langle \omega \partial _p s \rangle =  \langle R \rangle + H,
\label{eq:primeq_T''}\\
\langle \omega \partial _p q \rangle =  E,
\label{eq:primeq_q''}
\end{eqnarray}
where $H$ and $E$ are the surface sensible and latent heat fluxes, and $\langle \rangle$ indicates the integral from the surface to the top of the inversion $p_{\rm inv}$.

In our formulation, $\omega$ varies linearly from $p_{\rm inv}$ to the surface. Equations (\ref{eq:primeq_T''}) and (\ref{eq:primeq_q''}) can therefore be rewritten:
\begin{eqnarray}
\omega_{\rm inv} (s_M - s^+_{\rm inv})  =  \langle R \rangle + H,
\label{eq:primeq_T'''}\\
\omega_{\rm inv} q_M  =  E,
\label{eq:primeq_q'''}
\end{eqnarray}
where the subscript $M$ indicates the vertically averaged value over the ABL and $s^+_{\rm inv}$ is the dry static energy at the top of the inversion. We used the fact that the humidity at that level is zero.

Equation (\ref{eq:primeq_q'''}) gives us one constraint that must be satisfied
in order for the dry solution to occur: the ventilation of the ABL by the subsidence has to compensate the evaporation. On the other hand, Equation (\ref{eq:primeq_T'''}) shows that the warming by surface sensible heat flux and subsidence has to be compensated by the radiative cooling. The two equations can be rewritten as a single condition for the existence of the non-convective equilibrium:
\beq
- \frac{\langle R \rangle + H}{s^+_{\rm inv} - s_M} \; q_M = E.
\label{eq:ominv}
\eeq 
As the SST increases, $s_M$ increases, while $s^+_{\rm inv}$ is fixed in WTG. The ABL stability $s^+_{\rm inv} - s_M$ thus decreases. The radiative cooling $- \langle R \rangle $ increases too, and the surface heat flux $H$ is not very sensitive to SST changes. The subsidence $\omega_{\rm inv}$ therefore must increase with increasing SST. The ABL specific humidity $q_M$ also increases with SST. So the effect of ventilation increases with SST. 

To maintain the dry solution over an ocean surface, it is important that the ABL air be able to stay relatively moist as the SST increases so that $E$ is limited and the ventilation $\omega_{\rm inv} q_M$ is efficient. Otherwise $E$ will become very large due to the large air-sea humidity contrast (taking surface wind fixed), and in general there is no mechanism for $R$ nor $(s^+_{\rm inv} - s_M)^{-1}$ to become large at the same time to compensate.  
In nature (and in many models), the ABL generally does stay moist under a dry free troposphere, e.g., in subtropical trade wind regions.  In some models, though, this may not occur;  obvious examples are idealized models in which the shape of the entire vertical structure of the humidity field is fixed, so that the
ABL moisture is proportional to the free-tropospheric moisture. 
This happens,
for example, in the first quasi-equilibrium 
tropical circulation model (QTCM;  \cite{NZ00,ZNC00}).  This can be
remedied by allowing a separate
degree of freedom for boundary layer moisture, as in some otherwise similar
models \citep{WL93,Neggers06,SN06,KM06}.

The second constraint that must hold in order for the dry solution to occur
is that the convective heating and moistening remain zero, or at least small.
For a dry free troposphere and a cold sea surface, so that CAPE is negative,
this will occur for any reasonable convective parameterization.  For a dry
free troposphere and a warm sea surface, so that non-entraining
CAPE is significantly positive,
the results may be model-dependent.  Parameterizations which are insufficiently
sensitive to dry free tropospheric air may be able to generate some heating.
Once heating occurs, if $Q_c + R > 0$ there will be ascent, which will moisten
the troposphere, leading to the establishment of the rainy solution.
Alternatively, if the scheme is not able to generate deep convection and
associated heating, but can produce enough moistening above the ABL to 
eventually allow deep convection to occur, this will also lead to the
rainy solution.  It is reasonable to expect that under the same boundary
conditions, some models will have multiple equilibria and others will not.

We have performed preliminary experiments with a small number of other
SCMs in addition to the GEOS5.  In these experiments, we have been able
to produce multiple equilibria in some but not all of the models.  However,
our search over the space of initial and boundary conditions in these
other models has been cursory, so it is premature to draw any conclusions
from these results.  We will report on them in more detail when we have
performed a more thorough study.

\section{Conclusions}

We have shown that a single-column model using essentially the same
physics and numerics as a state-of-the-art GCM has multiple equilibria,
when run with fixed free tropospheric temperature and diagnostic 
large-scale vertical velocity according to the weak temperature 
gradient approximation.  When the boundary conditions are such as to
allow a rainy equilibrium state, a second equilibrium with no
precipitation and a very dry free troposphere also exists, and is
reached by initializing the model with a very dry sounding.  This
dry equilibrium state persists in the presence of parameterized 
horizontal moisture advection, represented as a relaxation of
the specific humidity back to a relatively moist reference profile, 
as long as the relaxation time scale is not too short.  The critical 
value of this time scale for existence of multiple equlibria is about four 
days in this model.  When the SST is sufficiently low, compared to 
that at which the free-tropospheric temperature profile 
would be in radiative-convective equilibrium (RCE),
only the dry equilibrium exists.

The existence of the dry equilibrium under SST greater than the 
RCE value requires that the radiatively-driven
large-scale descent be able to export sufficient moisture to 
balance the surface evaporation, and that the convective
parameterization (or explicit convection, in a cloud-resolving
model) be sufficiently sensitive to free-tropospheric moisture
that the dry troposphere inhibits deep convection from occurring.
Whether these requirements are met may be model-dependent.

The existence of these multiple equilibria is a consequence
of the interaction between deep convection and large-scale
dynamics, with the latter parameterized here through the weak
temperature gradient approximation.  The multiple equilibria
presented here are in this respect fundamentally different than
those found by \citet{Renno97}, whose model did not include a
representation of large-scale dynamics.  It seems possible, but is
not obvious, that analogous multiple equilibria would exist for
other parameterizations of large scale dynamics, similar in
spirit to the WTG approach used here but differing in detail
(e.g., \citet{Bergman_Sardeshmukh_04,Mapes04,Kuang07}).

The existence of these multiple equilibria strikes us as a
direct and simple expression of the tropical atmosphere's 
inherent nonlinearity.  It illustrates the complexity that is hidden behind the
averaging when the time-mean precipitation, for example, is 
considered as a function of boundary conditions.  

We cannot be certain that the multiple
equilibria would exist for a hypothetical model with ``correct'' physics.
Recent studies suggest, however, that the tendency of many
current convective parameterizations 
is to have too little, rather than too much, 
sensitivity to free tropospheric moisture, due to insufficient
entrainment (e.g., \citet{Derbyshire04,KB06,Biasutti06}).  As we expect that 
more sensitivity will make the multiple equilibria more likely
to exist, this suggests that they are not purely an
artifact of a bias in the GEOS5 physics.  In addition, in recent simulations
of radiative-convective equilibrium on a large domain over uniform SST 
\citep{Bretherton05} 
deep convection occurs only in a single, small region, while the rest of
the domain becomes extremely dry.  This behavior is analogous to what
we find here, with our single column representing either the dry or rainy
region separately.  This suggests that a cloud-resolving model run in WTG
mode might well exhibit multiple equilibria.

\begin{acknowledgments}
We thank David Neelin for a very helpful discussion. 
This work was supported by NASA grant NNX06AB48G S01.
\end{acknowledgments}


\begin{thebibliography}{15}
\providecommand{\natexlab}[1]{#1}
\expandafter\ifx\csname urlstyle\endcsname\relax
  \providecommand{\doi}[1]{doi:\discretionary{}{}{}#1}\else
  \providecommand{\doi}{doi:\discretionary{}{}{}\begingroup
  \urlstyle{rm}\Url}\fi

\bibitem[{\textit{Bergman and Sardeshmukh}(2004)}]{Bergman_Sardeshmukh_04}
Bergman, J.~W., and P.~D. Sardeshmukh (2004), Dynamic stabilization of single
  column models, \textit{Journal of Climate}, \textit{17}, 1004--1021.

\bibitem[{\textit{Biasutti et~al.}(2006)\textit{Biasutti, Sobel, and
  Kushnir}}]{Biasutti06}
Biasutti, M., A.~H. Sobel, and Y.~Kushnir (2006), {GCM} precipitation biases in
  the tropical {Atlantic}, \textit{Journal of Climate}, \textit{19}, 935--958.

\bibitem[{\textit{Bretherton et~al.}(2005)\textit{Bretherton, Blossey, and
  Khairoutdinov}}]{Bretherton05}
Bretherton, C.~S., P.~N. Blossey, and M.~Khairoutdinov (2005), An
  energy-balance analysis of deep convective self-aggregation above uniform
  {SST}, \textit{Journal of the Atmospheric Sciences}, \textit{62}, 4273--4292.

\bibitem[{\textit{Derbyshire et~al.}(2004)\textit{Derbyshire, Beau, Bechtold,
  Grandpeix, Piriou, Redelsperger, and Soares}}]{Derbyshire04}
Derbyshire, S.~H., I.~Beau, P.~Bechtold, J.-Y. Grandpeix, J.-M. Piriou, J.-L.
  Redelsperger, and P.~M.~M. Soares (2004), Sensitivity of moist convection to
  environmental humidity, \textit{Quarterly Journal of the Royal Meteorological
  Society}, \textit{130}, 3055--3080.

\bibitem[{\textit{Khouider and Majda}(2006)}]{KM06}
Khouider, B., and A.~J. Majda (2006), Multicloud convective parameterizations
  with crude vertical structure, \textit{Theor. Comp. Fluid Dyn.}, \textit{20},
  351--375.

\bibitem[{\textit{Kuang}(2007)}]{Kuang07}
Kuang, Z. (2007), Modeling the interaction between cumulus convection and
  linear waves using a limited domain cloud system resolving model,
  \textit{Journal of the Atmospheric Sciences}, \textit{in press}.

\bibitem[{\textit{Kuang and Bretherton}(2006)}]{KB06}
Kuang, Z., and C.~S. Bretherton (2006), A mass flux scheme view of a
  high-resolution simulation of a transition from shallow to deep cumulus
  convection, \textit{Journal of the Atmospheric Sciences}, \textit{63},
  1895--1909.

\bibitem[{\textit{Mapes}(2004)}]{Mapes04}
Mapes, B.~E. (2004), Sensitivities of cumulus-ensemble rainfall in a
  cloud-resolving model with parameterized large-scale dynamics,
  \textit{Journal of the Atmospheric Sciences}, \textit{61}, 2308--2317.

\bibitem[{\textit{Neelin and Zeng}(2000)}]{NZ00}
Neelin, J.~D., and N.~Zeng (2000), A quasi-equilibrium tropical circulation
  model - formulation, \textit{Journal of the Atmospheric Sciences},
  \textit{57}, 1741--1766.

\bibitem[{\textit{Neggers et~al.}(2006)\textit{Neggers, Stevens, and
  Neelin}}]{Neggers06}
Neggers, R., B.~Stevens, and J.~D. Neelin (2006), A simple equilibrium model
  for shallow cumulus convection, \textit{Theor. Comp. Fluid Dyn.},
  \textit{20}, 305--322.

\bibitem[{\textit{Raymond and Zeng}(2005)}]{RZ05}
Raymond, D.~J., and X.~Zeng (2005), Modeling tropical convection in the context
  of the weak temperature gradient approximation, \textit{Quarterly Journal of
  the Royal Meteorological Society}, \textit{131}, 1301--1320.

\bibitem[{\textit{Renn\'{o}}(1997)}]{Renno97}
Renn\'{o}, N.~O. (1997), Multiple equilibria in radiative-convective
  atmospheres, \textit{Tellus}, \textit{49}, 423--438.

\bibitem[{\textit{Renn\'{o} et~al.}(1994)\textit{Renn\'{o}, Emanuel, and
  Stone}}]{Renno94a}
Renn\'{o}, N.~O., K.~A. Emanuel, and P.~H. Stone (1994), Radiative-convective
  model with an explicit hydrologic cycle. 1: Formulation and sensitivity to
  model parameters, \textit{Journal of Geophysical Research}, \textit{99},
  14,429--14,441.
  
\bibitem[{\textit{Sobel and Bretherton}(2000)}]{SB00}
Sobel, A.~H., and C.~S. Bretherton (2000), Modeling tropical precipitation in a
  single column, \textit{Journal of Climate}, \textit{13}, 4378--4392.

\bibitem[{\textit{Sobel and Neelin}(2006)}]{SN06}
Sobel, A.~H., and J.~D. Neelin (2006), The boundary layer contribution to
  intertropical convergence zones in the quasi-equilibrium tropical circulation
  model framework, \textit{Theoretical and Computational Fluid Dynamics},
  \textit{20}, 323--350.

\bibitem[{\textit{Wang and Li}(1993)}]{WL93}
Wang, B., and T.~Li (1993), A simple tropical atmosphere model of relevance to
  short-term climate variations, \textit{Journal of the Atmospheric Sciences},
  \textit{50}, 260--284.

\bibitem[{\textit{Zeng et~al.}(2000)\textit{Zeng, Neelin, and Chou}}]{ZNC00}
Zeng, N., J.~D. Neelin, and C.~Chou (2000), A quasi-equilibrium tropical
  circulation model - implementation and simulation, \textit{Journal of the
  Atmospheric Sciences}, \textit{57}, 1767--1796.

\end{thebibliography}



\clearpage

\clearpage

\clearpage
\begin{figure}
\noindent
\includegraphics[height=6cm,clip]{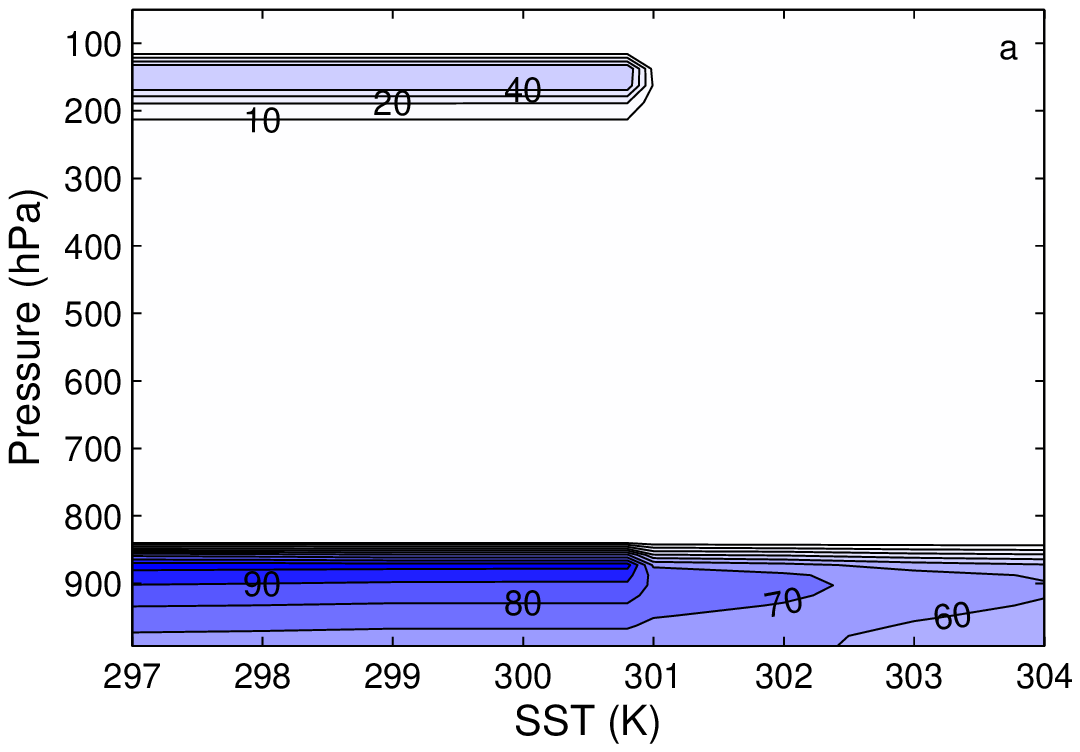}
\includegraphics[height=6cm,clip]{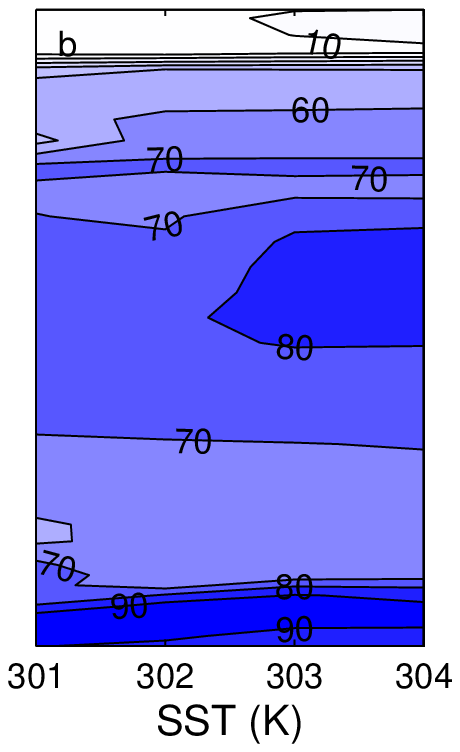}
\caption{Relative humidity (\%) as a function of SST and pressure in the non-convective (a) and convective (b) equilibria, with no horizontal moisture advection.}
\label{fig:rh}
\end{figure}
\clearpage
\end{article}
\end{document}